\documentclass[12pt]{iopart} 
\usepackage{epsfig,amssymb}  
\begin{document} 
\title[]{Determining the WIMP mass using direct detection experiments} 
\author{Anne M.~Green\dag}
\address{\dag\ School of
Physics and Astronomy, University of Nottingham, University Park,
  Nottingham, NG7 2RD, UK}
\eads{\mailto{anne.green@nottingham.ac.uk}}, 
\begin{abstract} 

We study the accuracy with which the WIMP mass could be determined 
by a superCDMS-like direct
detection experiment, given optimistic assumptions about
the detector set-up and WIMP properties.
We consider WIMPs with an interaction
cross-section of $\sigma_{\rm p} = 10^{-7}\, {\rm pb}$ (just below current
exclusion limits) and assume, initially, 
that the local WIMP velocity distribution
and density are known and that the experiment has negligible background.
For light WIMPs (mass significantly less than that of the target nuclei) small
variations in the WIMP mass lead to significant changes in the energy
spectrum.  Conversely for heavy WIMPs 
the energy spectrum depends only weakly on the WIMP
mass. Consequently it will be far easier to measure the WIMP mass if
it is light than if it is heavy. 
With exposures of ${\cal E}= 3 \times 10^{3}, \, 
3 \times 10^{4} \,$ and $3 \times 10^{5} \, {\rm kg \, day}$
(corresponding, roughly, to the three proposed
phases of SuperCDMS) it will be possible,
given the optimistic assumptions mentioned above, 
to measure the mass of a light
WIMP with an accuracy of roughly $25\%, 15\%$ and $2.5 \%$ respectively.
These numbers increase with increasing WIMP mass, and for heavy WIMPs,
$m_{\chi} > {\cal O}(500 \, {\rm GeV})$, even with a large exposure it will
only be possible to place a lower limit on the mass.
Finally we discuss the validity of the various assumptions made,  and the 
consequences if these assumptions are not valid.
In particular if the local WIMP distribution is composed
of a number of discrete streams it will not be possible
to determine the WIMP mass.
\end{abstract}

%  JCAP uses keywords rather than pacs numbers. 
 
\begin{flushleft} {\bf Keywords}: dark matter, dark matter detectors
\end{flushleft} 
 
 
\maketitle 
 
\section{Introduction} 
\label{intro} 
 
Diverse cosmological observations indicate that the majority of the
matter in the Universe is dark and
non-baryonic (e.g. Ref.~\cite{cosmo}).  Weakly Interacting Massive
Particles (WIMPs) are one of the leading cold dark matter candidates,
and supersymmetry provides a concrete, well-motivated, WIMP candidate
in the form of the lightest neutralino (e.g. Ref.~\cite{jkg,dmrev}).
The direct detection of WIMPs in the lab~\cite{ddtheory} 
would not only directly
confirm the existence of dark matter but would also probe the parameters
of supersymmetry models. Constraints on, or measurements of, the WIMP
mass and interaction cross-section will be
complementary to the information derived from collider 
experiments~\cite{collider1,collider2}. 
It is therefore pertinent to examine the accuracy with which WIMP
direct detection experiments will be able to measure the WIMP mass, if
they detect  WIMPs. 

Direct detection experiments could potentially measure the WIMP mass
via the mass dependence of either the energy spectrum~\cite{ls} or
the `crossing energy' at which the phase of the annual modulation
signal~\cite{amtheory}, due to the motion of the Earth, changes
sign~\cite{amphase,hasenbalg,lf}.  The size of the annual
modulation signal is small (of order a few per-cent), however, and given the
current exclusion limits from CDMS~\cite{CDMS1,CDMS2} it is unlikely that
even planned tonne scale experiments, such as
SuperCDMS~\cite{SuperCDMS1,SuperCDMS2}, Xenon~\cite{xenon} and
EURECA~\cite{eureca}, will be able to accurately measure
the energy dependence of the annual
modulation phase~\cite{hasenbalg,ramachers}~\footnote{The DAMA collaboration
have, with a NaI detector and an exposure of
$\sim 1.1 \times \, 10^{5} \, {\rm kg \, day}$, observed an annual
modulation, which they interpret as a WIMP signal~\cite{dama}, however
it appears to be possible to reconcile this with the CDMS exclusion limit
only by invoking `non-standard' WIMP properties (such as the
WIMP-proton and WIMP-neutron couplings being different~\cite{kk} or 
WIMPs which can scatter inelastically~\cite{swin})
.}. Furthermore the shape, phase and amplitude of the annual modulation signal
depend sensitively on the detailed local WIMP velocity
distribution~\cite{amnonstand} whereas the shape of the
differential event rate does not~\cite{drdens,lgegreen}, 
if the local WIMP distribution is smooth. The mass dependence
of the differential event rate~\cite{ls} (see also~\cite{bk})
therefore appears to offer the best prospect, in the
short to medium term at least, for probing the WIMP
mass using direct detection experiments.

In this paper we examine the accuracy with which a future
SuperCDMS~\cite{SuperCDMS1,SuperCDMS2} like direct detection
experiment will be able to measure the WIMP mass, given a positive
detection.  We consider the effect of varying the underlying WIMP mass
and detector exposure, and also examine the uncertainties which arise
from our lack of knowledge of the underlying WIMP 
distribution. In Sec.~\ref{der} we outline the calculation
of the differential event rate and its dependence on the WIMP mass.  
We describe our Monte Carlo simulations and results in Sec.~\ref{MC},
discuss the validity of the assumptions made in Sec.~\ref{assumpt}
and conclude with discussion in Sec.~\ref{discuss}.

\section{Differential event rate} 
\label{der}

\subsection{Basic calculation}

The direct detection differential event rate, or energy spectrum,
depends on the WIMP mass, its interaction with the detector nuclei and
the velocity distribution of the incoming WIMPs. Assuming purely
spin-independent coupling, the event rate per unit energy, 
is given by (see e.g.~\cite{jkg,ls}):
\begin{equation}
\label{drde}
\frac{{\rm d} R}{{\rm d}E}(E) =
             \frac{\sigma_{{\rm p}} 
             \rho_{\chi}}{2 \mu_{{\rm p} \chi}^2 m_{\chi}}
             A^2 F^2(E) {\cal F}(E)   \,, 
\end{equation}
where $\rho_{\chi}$ is the local WIMP density, $\sigma_{{\rm
p}}$ the WIMP scattering cross section on the proton, $\mu_{{\rm p} \chi} = 
(m_{\rm p} m_{\chi})/(m_{{\rm p}}+ m_{{\chi}})$ the WIMP-proton reduced mass, 
$A$ and $F(E)$ the mass number and
form factor of the target nuclei respectively and $E$ is the recoil
energy of the detector nucleus.
The dependence  on the WIMP velocity distribution is encoded in
${\cal F}(E)$, which is defined as
\begin{equation}
\label{tq}
{\cal F}(E)= \langle \int^{\infty}_{v_{{\rm min}}} 
            \frac{f^{\rm E}(v,t)}{v} {\rm d}v  \rangle \,,
\end{equation}
where $f^{\rm E}(v,t)$ is the (time dependent) 
WIMP speed distribution in the rest frame of the
detector, normalized to unity and $\langle .. \rangle$ 
denotes time averaging. This is calculated from the
velocity distribution in the rest frame of the Galaxy,
$f^{\rm G}({\bf v})$, via Galilean transformation:
${\bf v} \rightarrow \tilde{{\bf v}}= {\bf v} + {\bf v}^{\rm E}(t)$
where ${\bf v}^{\rm E}(t)$ is the Earth's velocity with
respect to the Galactic rest frame.
The lower limit of the integral, $v_{{\rm min}}$, is the minimum
WIMP speed that can cause a recoil of energy $E$:
\begin{equation}
\label{vmin}
v_{{\rm min}}=\left( \frac{ E m_{A}}{2 \mu_{{\rm A} \chi}^2} 
             \right)^{1/2} \,,
\end{equation}
where $m_{A}$ is the atomic mass of the detector nuclei
and $\mu_{{\rm A} \chi}$ the WIMP-nucleon reduced mass.
We use the Helm form factor~\cite{helm} with parameter values as 
advocated by Lewin and Smith~\cite{ls}. For most calculations we 
use  the `standard halo model',
an isotropic isothermal sphere, for which
 the local WIMP velocity distribution, in the Galactic
rest frame, is Maxwellian (c.f. Ref.~\cite{amtheory})
\begin{eqnarray}
f^{\rm G}({\bf v}) & = & N \left[ \exp{\left(- 
    |{\bf v}|^2/v_{\rm c}^2 \right)} -
         \exp{\left(- v_{\rm esc}^2/v_{\rm c}^2 \right)} \right] 
     \hspace{1.0cm} |{\bf v}|< v_{\rm esc} \,, \\
f^{\rm G}({\bf v}) & = & 0  \hspace{7.0cm} |{\bf v}|> v_{\rm esc} \,,
\end{eqnarray}
where $N$ is a normalization factor, $v_{\rm c}=220 \, 
{\rm km \, s}^{-1}$~\cite{klb}
 and $v_{\rm esc}=540 \, {\rm km \, s}^{-1}$~\cite{rave}
 are the local circular and escape speeds respectively
and we use the usual fiducial value for the
local WIMP density,  $\rho_{\chi} =0.3 \, {\rm GeV \, cm}^{-3}$.
We use the expressions for the time dependence of the Earth's velocity 
with respect to the Galactic rest frame from Ref.~\cite{ls},
to transform the velocity distribution into the lab frame.

\subsection{Dependence on the WIMP mass}

\begin{figure} 
\begin{center} 
\epsfxsize=5.in 
\epsfbox{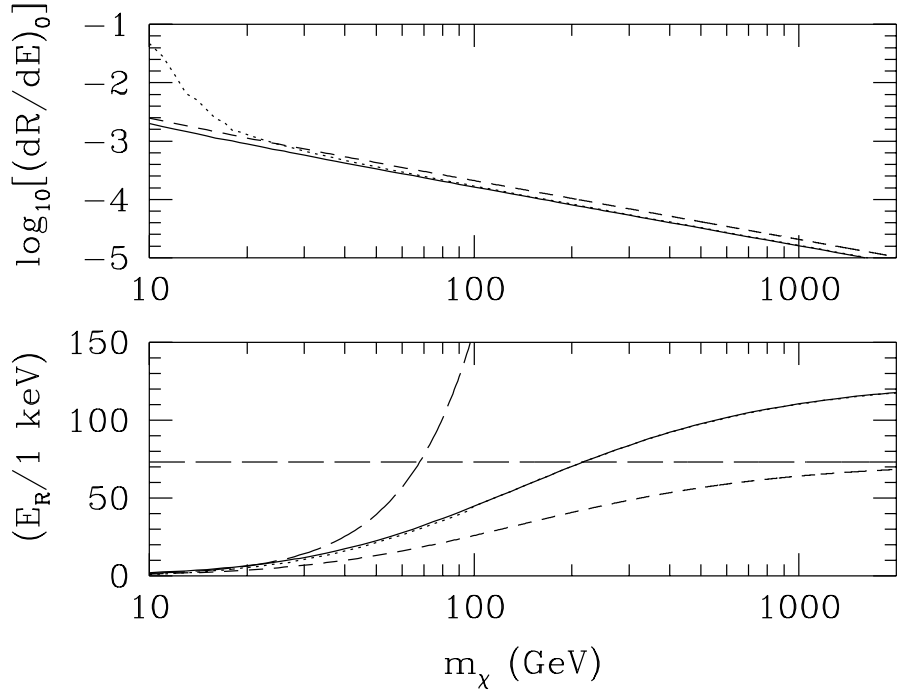}
\end{center}
\caption{
The differential event rate in the $E \rightarrow 0\, {\rm keV}$ limit,
$({\rm d} R/{\rm d} E)_0$, (top panel)
and the characteristic energy scale, $E_{\rm R}$, (bottom panel)
as a function of the WIMP mass, $m_{\chi}$ for a ${\rm Ge}$ detector. 
The solid (dotted) lines are the fits to the 
full calculation including the Earth's velocity and the Galactic
escape speed for threshold energy
$E_{\rm th}=0 \, (10) \, {\rm keV}$. The short dashed lines
are eqs.~(\ref{pre}) and (\ref{er}) which neglect the
Earth's velocity and the Galactic escape speed. In the bottom panel
the long dashed lines are the asymptotic mass dependences of $E_{\rm R}$
in the $m_{\chi} \ll  m_{\rm A}$ and $\gg  m_{\rm A}$ limits
($E_{\rm R} \propto m_{\chi}^2$ and $\propto {\rm const}$ respectively).
\label{ab}}
\end{figure}

The WIMP mass dependence of the differential event rate can most
easily be seen, following Lewin and Smith~\cite{ls}, by first
neglecting
the Earth's velocity and the Galactic escape speed. In this case
eq.~(\ref{drde}) can be written as
\begin{equation}
\label{drde0}
\frac{{\rm d} R}{{\rm d}E}(E) = \left(\frac{{\rm d} R}{{\rm d}E}\right)_{0}
              \exp{ \left( -\frac{E}{E_{\rm R}} \right)}
              F^2(E)  \,,  
\end{equation}
where $({\rm d} R/{\rm d}E)_{0}$, the event
rate in the $E \rightarrow 0\, {\rm keV
}$ limit, and $E_{\rm R}$, the characteristic
energy scale, are given by
\begin{equation}
\label{pre}
\left(\frac{{\rm d} R}{{\rm d}E}\right)_{0}= \frac{\sigma_{\rm p} \rho_{\chi}}
         {\sqrt{\pi} \mu_{{\rm p} \chi} ^2 m_{\chi} v_{c}} A^2 \,,
\end{equation}
and
\begin{equation}
\label{er}
E_{\rm R} = \frac{2 \mu_{{\rm A} \chi}^2 v_{c}^2}{m_{\rm A}} \,,
\end{equation}
respectively.
When the Earth's velocity and the Galactic escape speed
are taken into account eq.~(\ref{drde}) is still a reasonable
approximation to the event rate, provided that multiplicative constants  
$c_{0}$ and $c_{E_{\rm R}}$, are included in the expressions for
$\left(\frac{{\rm d} R}{{\rm d}E}\right)_{0}$ and $E_{\rm R}$~\cite{ls}.
The exact values of the constants, which are of order unity,
depend on the target nuclei, the energy threshold and the Galactic
escape speed.
We find, using least squares fitting to the full numerically calculated
time averaged differential event for a ${\rm Ge}$ detector
with energy threshold $E_{\rm th}=0 \, {\rm keV}$, $c_{0} \approx 0.78$
and $c_{E_{\rm R}} \approx 1.72$, with a weak dependence 
(in the 3rd significant
figure) on the WIMP mass.

The mass dependence of $({\rm d} R/{\rm d}E)_{0}$
and $E_{\rm R}$ for a ${\rm Ge}$ detector are shown in fig.~\ref{ab}. 
As expected from
eq.~(\ref{pre}), $({\rm d} R/{\rm d}E)_{0}
\propto m_{\chi}^{-1}$, since $m_{\chi} \gg m_{\rm p}$. 
Similarly as expected from eq.~(\ref{er}),
for $m_{\chi} < m_{\rm A}$, 
$E_{\rm R} \propto m_{\chi}^{2}$. This mass dependence weakens with 
increasing WIMP mass, and $E_{\rm R}$
tends to its constant large mass asymptote for $m_{\chi} \sim {\cal O}( 1
\, {\rm TeV})$. We will see below that this has important consequences
for the determination of the WIMP mass from direct detection
experiments.  If, anticipating section III, we only fit
to the event rate above an energy threshold $E_{\rm th}=10\, {\rm keV}$
we find that for WIMP masses below 
$ \sim 100 \, {\rm GeV}$ the fitting constants
depend on the WIMP mass (this is visible in fig.~\ref{ab} as the
deviation of the dotted lines from the solid lines for small WIMP masses). 
This illustrates that while  eq.~\ref{drde0}
is a useful approximation
for demonstrating the mass dependence of the energy spectrum,
for concrete applications the time average of the local WIMP velocity 
distribution should be calculated explicitly.

\section{Monte Carlo simulations}
\label{MC} 
We use Monte Carlo simulations to examine, for a range of detector
exposures and input WIMP masses, how well the WIMP mass could be
determined from the energies of observed WIMP nuclear recoil events.

\subsection{Detector properties}
Our simulated detector is based on the proposed SuperCDMS
experiment~\cite{SuperCDMS1,SuperCDMS2}, being composed of Ge with a
nuclear recoil energy threshold $E_{\rm th}=10 \, {\rm keV}$. We assume that
the background event rate is negligible, as is expected for this
experiment located at SNOLab~\cite{SuperCDMS1,SuperCDMS2},
and that the energy resolution is perfect.  For
simplicity we assume that the nuclear recoil detection efficiency is
independent of energy. The energy dependence of the efficiency of the
current CDMS experiment is relatively small (it increases from $\sim
0.46$ at $E=20 \, {\rm keV}$, the energy above which data from all
detectors is analyzed, to $\sim 0.51$ at $ 100\, {\rm
keV}$~\cite{CDMS2}). For further discussion of these assumptions see
Sec.~\ref{assumpt}.

The proposed SuperCDMS detector consists of 3 phases, with detector
masses of $\sim 25 \, {\rm kg}, 150 \, {\rm kg}$ and $1 \, {\rm ton}$.
We consider efficiency weighted exposures~\footnote{For brevity
we subsequently refer to this as simply the exposure.}
${\cal E}=3 \times 10^{2}, 3
\times 10^{3}$, $3 \times 10^{4}$ and $3 \times 10^{5} \, {\rm kg \,
day}$. The later three exposures correspond, roughly, to a detector
with mass equal to that of the 3 proposed phases of SuperCDMS taking
data for a year~\footnote{Accumulating this much `live-time' would of
course take substantially longer than a year.} with a $\sim 50\%$
detection efficiency.

\subsection{WIMP properties}

We assume a fixed WIMP-nucleon cross-section of $\sigma_{\rm p}
=10^{-7} \, {\rm pb}$, which is just below the current exclusion limit
from the CDMS experiment~\cite{CDMS1}. Since the number of events
expected in a given experimental set-up is directly proportional to
the cross-section, there is a straight-forward scaling to other
cross-sections i.e. an exposure ${\cal E}=3 \times 10^{4}\,{\rm kg \,
day}$ for $\sigma_{\rm p}=10^{-7} \, {\rm pb} $ is equivalent to an
exposure of ${\cal E}=3 \times 10^{5}\, {\rm kg \, day}$ for
$\sigma_{\rm p}=10^{-8} \, {\rm pb}$.  We consider input WIMP masses of
$m_{\chi} = 25, 50, 100, 250$ and $500 \, {\rm GeV}$, with a Maxwellian
speed distribution with circular speed $v_{\rm c}=220 \, {\rm km
\, s}^{-1}$. For $m_{\chi}=100 \, {\rm GeV}$ we also consider circular
speeds in the range $v_{\rm c}=180$ to $260 {\rm km\,
s}^{-1}$~\cite{klb} and variations in the form of the speed
distribution.

\subsection{Statistical analysis} 

We estimate the WIMP mass and cross-section by maximizing the extended
maximum function (which takes into account the fact that the number of
events observed in a given experiment is not fixed)
e.g. Ref.~\cite{cowan}:
\begin{equation} 
  L= \frac{\lambda^{N_{\rm expt}} \exp{(-\lambda)}}{N_{\rm expt}!}
    \Pi_{i=1}^{N_{\rm expt}} f(E_{\rm i}) \,.
\end{equation}
Here $N_{\rm expt}$ is the number of events observed,
$E_{\rm i} \, (i=1,..., N_{\rm expt})$ are the energies
of the events observed, $f(E)$ is
the, normalized, differential event rate and
$\lambda = {\cal E} \int_{E_{th}}^{\infty} ({\rm d} R/{\rm d}
E) \, {\rm d} E$ is the mean number of events 
($f(E)$ and $\lambda$ depend on $m_{\chi}$ and $\sigma_{\rm p}$).
We calculate the probability distribution of the maximum likelihood estimators,
for each exposure and input WIMP mass, by simulating
$10^{4}$ experiments. We first calculate the expected number of
events, $\lambda_{\rm in} = {\cal E} \int_{E_{th}}^{\infty} ({\rm d} R/{\rm d}
E) \, {\rm d} E$, from the input energy spectrum. The actual number of
events for a given experiment, $N_{\rm expt}$, is drawn from a
Poisson distribution with mean $\lambda_{\rm in}$. We 
Monte Carlo generate $N_{\rm expt}$ events from the input energy spectrum, 
from which the maximum likelihood (hereafter `ML') mass and cross-section
are calculated.

\begin{figure}  
\begin{center}  
\epsfxsize=6.in  
\epsfbox{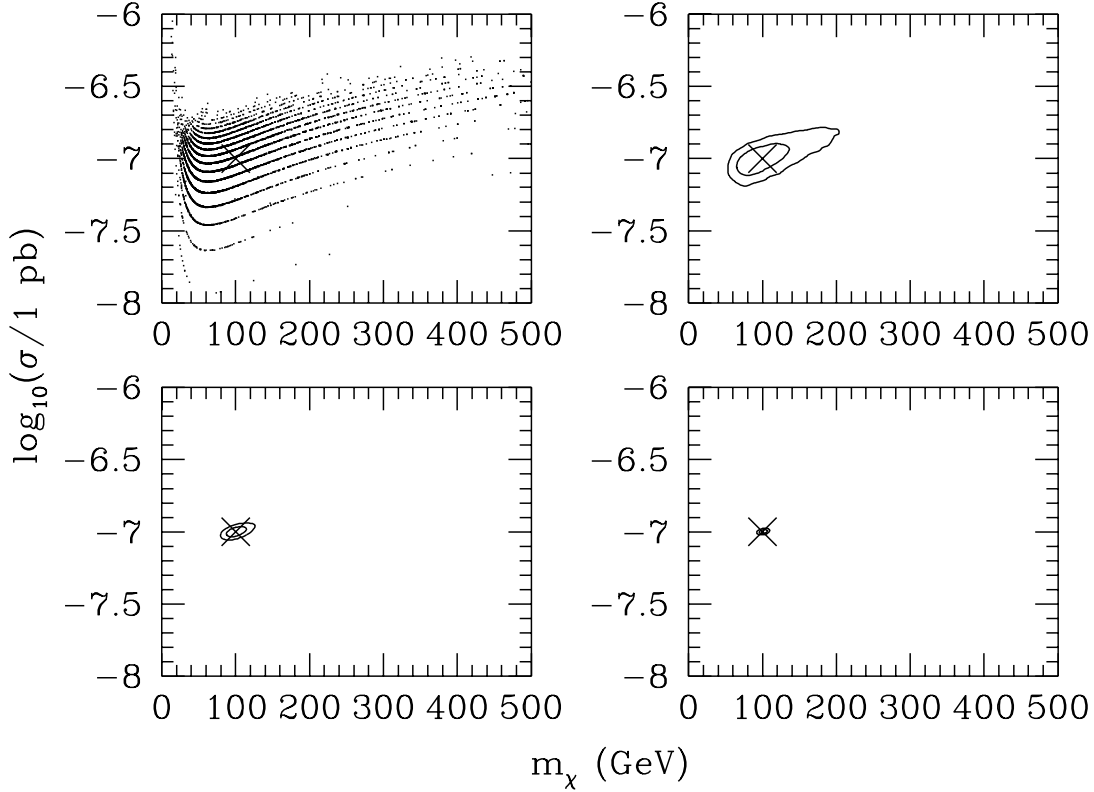} 
\end{center}  
\begin{center} 
\caption{The distribution of the maximum likelihood 
WIMP masses, $m_{\chi}$, and
cross-sections, $\sigma_{\rm p}$, for exposures of
(top row, left to right and then
bottom row left to right) 
${\cal E}=3 \times 10^{2},\,3 \times 10^{3},\,3 \times 10^{4}$ and
$ 3 \times 10^{5}\, {\rm kg \, day}$. For ${\cal E}= 3 \times 10^{2} \,
{\rm kg \,  day}$ we explicitly plot the results from all $10^{4}$ Monte Carlo
experiments. For the larger exposures we plot contours containing  
$68\%$ and $95\%$ of the probability distribution.
In each panel the large cross denotes the input parameters:
$m_{\chi}=100\, {\rm GeV}, \, \sigma_{\rm p}=10^{-7} \, {\rm pb}.$
\label{m100}}  
\end{center} 
\end{figure} 

\begin{figure}  
\begin{center}  
\epsfxsize=6.in  
\epsfbox{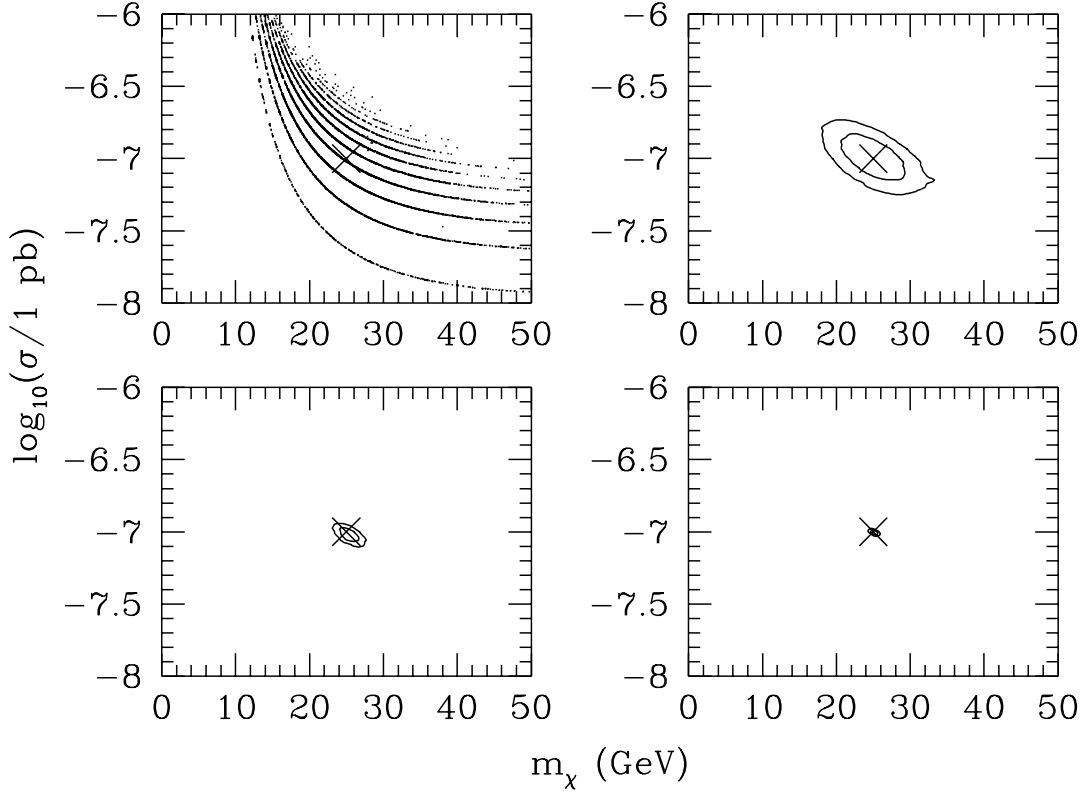} 
\end{center}  
\begin{center} 
\caption{As fig.~\ref{m100} for
$m_{\chi}=25\, {\rm GeV}$.
For ${\cal E}= 3 \times 10^{2} \,
{\rm kg \, day}$, $\lambda_{\rm in}=4.2$ and consequently there is a $\sim 1\%$
probability that an experiment would not detect any events.
\label{m25}}  
\end{center} 
\end{figure} 

\begin{figure}  
\begin{center}  
\epsfxsize=6.in  
\epsfbox{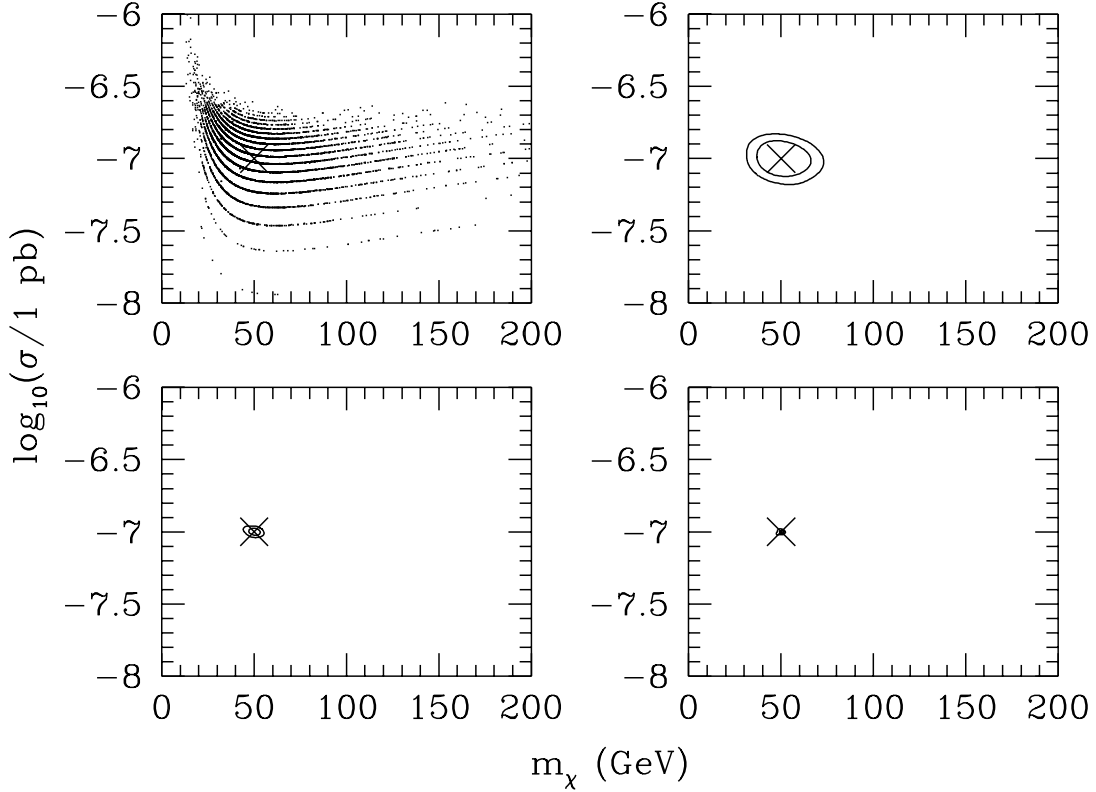} 
\end{center}  
\begin{center} 
\caption{As fig.~\ref{m100} for
$m_{\chi}=50\, {\rm GeV}$.
\label{m50}}  
\end{center} 
\end{figure} 

\begin{figure}  
\begin{center}  
\epsfxsize=6.in  
\epsfbox{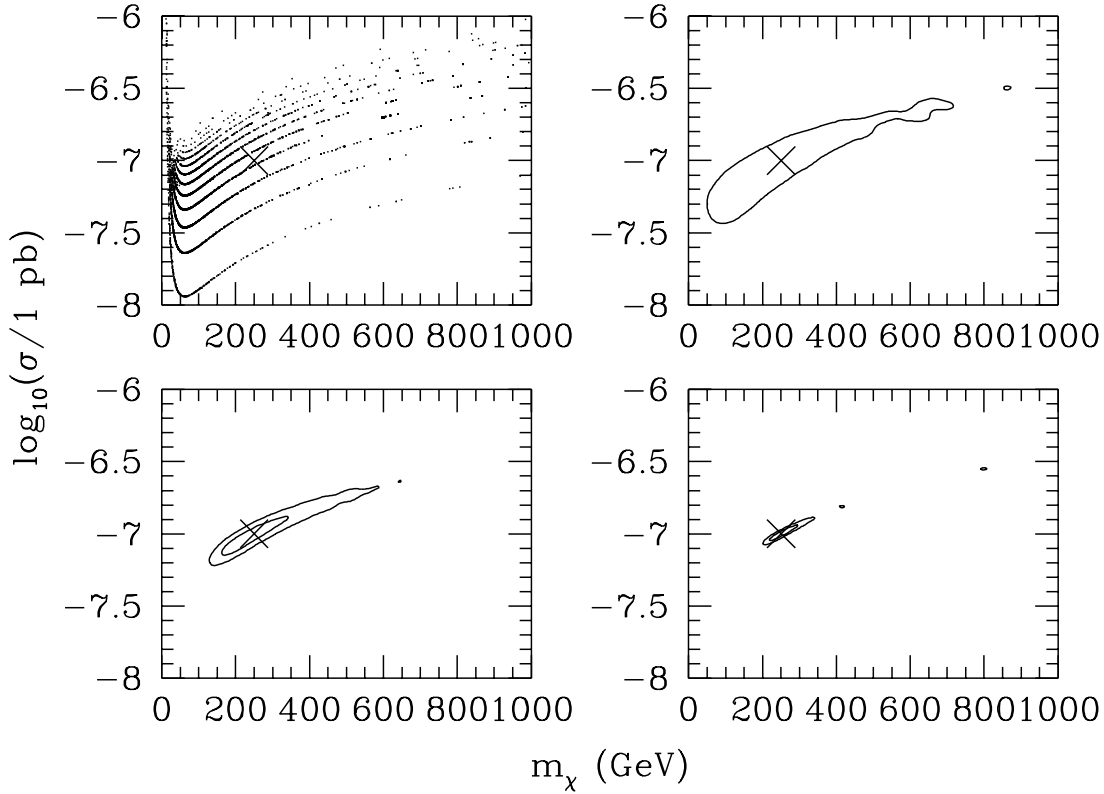} 
\end{center}  
\begin{center} 
\caption{As fig.~\ref{m100} for
$m_{\chi}=250\, {\rm GeV}$.
For ${\cal E}= 3 \times 10^{2} \,
{\rm kg \, day}$, $\lambda_{\rm in}=4.3$ and consequently there
is a $\sim 1\%$ probability that an
experiment would not detect any events. For ${\cal E}= 3 \times
10^{3} \, {\rm kg \, day}$ there is a low probability density tail extending
to large $m_{\chi}$, and hence we can not accurately calculate
a contour containing $95\%$ of the probability density. The disconnected
`blobs' at large $m_{\chi}$ are also a consequence of this extended 
low probability density tail.
\label{m250}}  
\end{center} 
\end{figure} 

\begin{figure}  
\begin{center}  
\epsfxsize=6.in  
\epsfbox{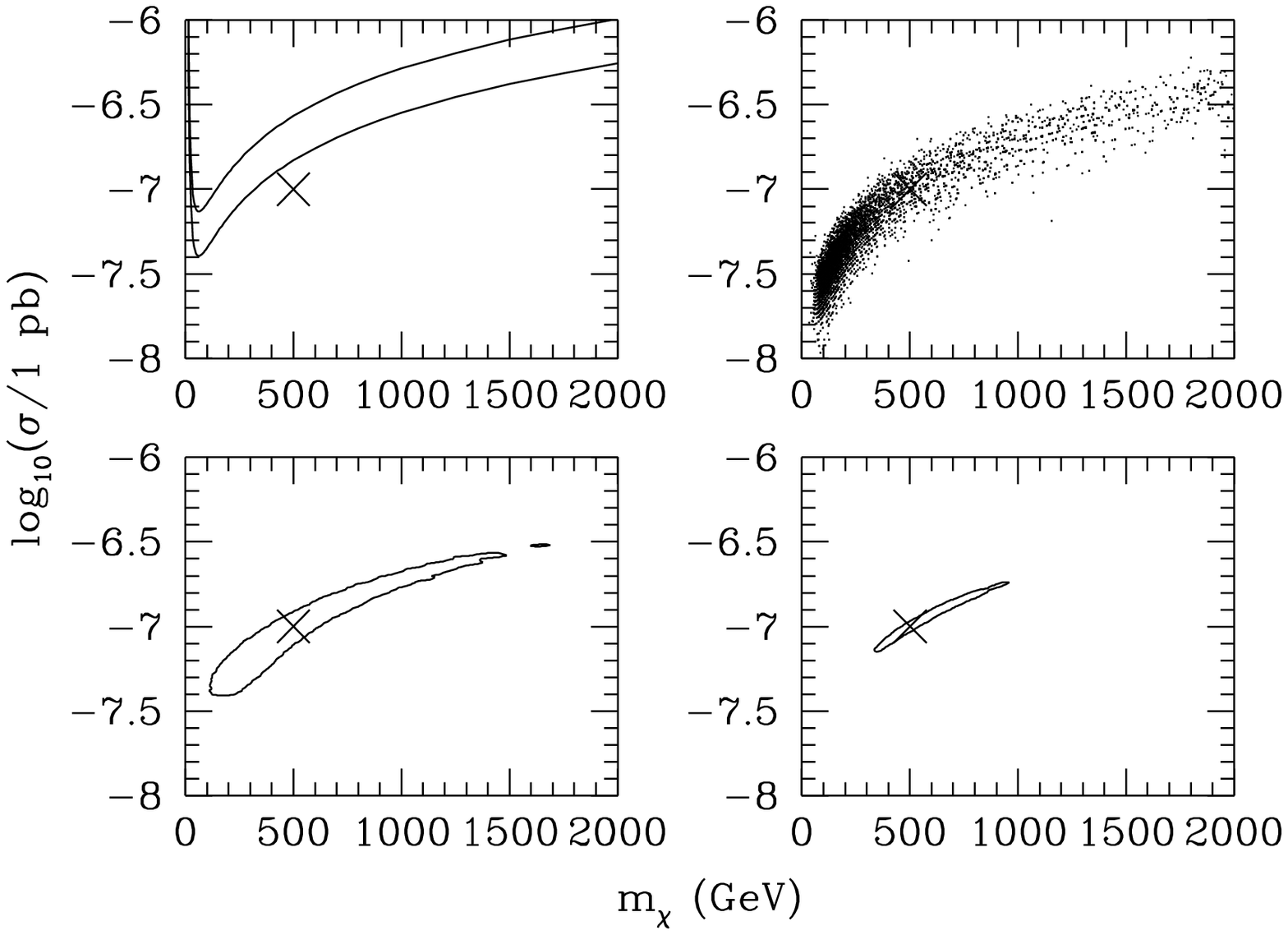} 
\end{center}  
\begin{center} 
\caption{As fig.~\ref{m100} for
$m_{\chi}=500\, {\rm GeV}$.
For ${\cal E}= 3 \times 10^{2} \,
{\rm kg \, day}$, $\lambda_{\rm in}=2.4$ and there is a $>50\%$
probability that an experiment
would observe two or fewer events.  Consequently we instead plot
the $68\%$ and $95\%$  exclusion limits which would be found
by an experiment observing $N_{\rm exp}=2$ events (the number of observed
events with the largest probability). For this input mass there is a low
probability density tail extending to large $m_{\chi}$, even
for ${\cal E}= 3 \times
10^{5} \, {\rm kg \, day}$, and hence it is not possible to plot
$95\%$ probability contours.
\label{m500}}  
\end{center} 
\end{figure} 

Finally we plot the distribution of ML  masses and cross 
sections. Given real data from a single experiment a
Bayesian analysis with priors on the WIMP parameters,
possibly based on the results of other experiments, would
be a reasonable approach.  However, the
question which we are trying to address (`For a given underlying WIMP
mass, how well can the mass be measured?')  is best answered by
simply considering the distribution
of ML masses and cross-sections.
Since the approximations we make regarding the detector set-up and
WIMP distribution and cross-section are optimistic, a real experiment
would make less accurate determinations of the WIMP properties.

\subsection{Results}

In fig.~\ref{m100} we plot the probability distribution of ML WIMP masses and
cross-sections for input WIMP parameters $m_{\chi}=100 \, {\rm GeV}$
and $\sigma_{\rm p}=10^{-7} \, {\rm pb}$. In this case both the input
energy spectrum and the maximum likelihood analysis of the simulated
events are carried out assuming a Maxwellian speed distribution with
$v_{\rm c}= 220 \, {\rm km \, s}^{-1}$. For each experiment the
extended likelihood is maximized for WIMP parameters which produce an
expected number of events equal to the actual number of
events observed in that experiment: $\lambda(m_{\chi}, \sigma_{\rm p})
= N_{\rm expt}$. This means that, for fixed exposure, the ML
parameters are localized on curves corresponding to fixed $N_{\rm
  expt}$. For a given experiment the position of the ML parameters on
the curve depends on the energies of the observed events. For ${\cal
  E}= 3 \times 10^{2} \, {\rm kg \, day}$, $\lambda_{\rm in}= 7.8$,
which is sufficiently small that the stratification of ML parameters
is clearly visible and we hence plot the actual pairs of
$m_{\chi}-\sigma_{\rm p}$ values. For the larger exposures the mean
number of events expected is proportionately larger, the
stratification is no longer visible, the ML values are better
localized in the $m_{\chi}-\sigma_{\rm p}$ plane and we instead plot
contours containing $68\%$ and $95\%$ of the simulated experiments. We
calculate the continuous probability distribution of $m_{\chi}$ and
$\sigma_{\rm p}$ by smoothing the ML values from the $10^{4}$ Monte
Carlo simulations with a double gaussian kernel and summing them. 
The deviations of the kernel are chosen, for each exposure and underlying mass,
to produce relatively smooth contours without artificially inflating the 
spread in the ML values. We then integrate the probability density above a 
threshold value and vary this threshold to find the values which enclose
$68\%$ and $95\%$ of the probability distribution, and plot contours
corresponding to these threshold values.

For the two smallest exposures there is a significant low probability
density tail at large ML masses. This is due to the weak mass
dependence of the characteristic energy $E_{\rm R}$ for $m_{\chi}>
m_{\rm A}$. For these exposures the majority of
experiments have ML masses smaller than the input mass. The
likelihood analysis is not biased, the expected value of the ML
mass is equal to the input mass, however there are a
significant number of experiments with ML masses substantially
larger than the input mass. As the exposure is increased,
$\lambda_{\rm in}$ becomes large and the fractional spread in the
number of events observed in each experiment becomes small. This
allows the interaction cross-section (which effectively acts as a
normalization factor) to be accurately determined. The larger number
of events also allows the energy dependence of the differential event
rate, and hence the mass, to be determined more accurately.

With an exposure of ${\cal E}=3 \times 10^{2} \, {\rm kg \, day}$ and
an underlying WIMP mass of $m_{\chi}=100 \, {\rm GeV}$
it will be difficult to make
quantitative statements about the WIMP mass, beyond excluding very
large or very small masses. Similarly it will not be possible to
measure the cross-section more accurately than to within an order
of magnitude. For ${\cal E}=3 \times 10^{3}
\, {\rm kg \, day}$ the accuracy with which the mass can be determined
improves, however there is still a significant tail of experiments
finding ML masses substantially larger than the underlying
value. More quantitatively $95\%$ of experiments will find ML
masses in the range $ 60 \, {\rm GeV} < m_{\chi} < 200 \, {\rm
GeV}$. Increasing the exposure by an order of magnitude, the
distribution of ML masses becomes more symmetric about the input
mass and the spread in the ML masses is of order $\pm 30 \,
{\rm GeV}$. With a further increase in the exposure to 
${\cal E}=3 \times 10^{5} \, {\rm kg \, day}$ the mass could be 
measured to an accuracy of around $10\, {\rm GeV}$ (see the top panel
of fig.~\ref{m100ns} for a zoom in on the distribution in this case).
The accuracy with which
the cross-section can be measured increases with increasing exposure: 
$\Delta (\log{\sigma_{\rm p}}) \sim \pm 0.2, 0.05$ and $0.02$ for ${\cal E}= 
3 \times 10^{3}, \, 3 \times 10^{4}$ and $3 \times 10^{5} \, {\rm kg\, day}$
respectively. 
The shape of the distribution of ML parameters reflects the
shape of curves of constant $N_{\rm expt}$ in the $m_{\chi} \sim 100 
\, {\rm GeV}$ region, with a weak positive correlation between the best-fit
values of $m_{\chi}$ and $\sigma_{\rm p}$.

In figs.~\ref{m25}-\ref{m500} we plot the results for input WIMP
masses of $m_{\chi}=25, \, 50,\, 250$ and $500\, {\rm GeV}$ respectively. For
$m_{\chi}=25, \, 250$ and $500\, {\rm GeV}$ the input mean number of
events for $E=3 \times 10^{2} \, {\rm kg \, day}$ is small enough that there
is a significant probability that an experiment will see no events (making it
impossible to determine the WIMP mass).  

For light WIMPs, $m_{\chi} < m_{\rm A}$, the characteristic energy,
$E_{\rm R}$, varies significantly with WIMP mass. This allows the 
mass to be determined with large exposures more accurately than for
$m_{\chi} = 100\, {\rm GeV}$. For an underlying 
mass of $m_{\chi} = 25 \, {\rm GeV}$ and an exposure of ${\cal E}= 2
\times 10^{3} \, {\rm kg \,day}$, due to the small expected number of
events, it will be difficult to place meaningful constraints (beyond
an upper limit on the mass) on the WIMP parameters. With larger
exposures it will be possible to measure the WIMP mass and
cross-section with increasing accuracy; for ${\cal E}= 3 \times 10^{3},
\, 3 \times 10^{4}$ and $3\times 10^{5} \, {\rm kg\, day}$ the
distribution of ML WIMP masses is symmetric and $95\%$ of
experiments lie within $\pm 6, 4$ and $1 \, {\rm GeV}$ of the input
mass respectively. The accuracy with which $\sigma_{\rm p}$ can be
measured improves roughly as for $m_{\chi}=100 \, {\rm GeV}$.
For $m_{\chi} \sim 50\, {\rm GeV}$ the fractional accuracy
with which $m_{\chi}$ can be measured is similar; $95\%$ of
experiments lie within $\pm 12, 7, 2 \, {\rm GeV}$ of the input mass
for ${\cal E}= 3
\times 10^{3}, \, 3 \times 10^{4}$ and $3 \times 10^{5} \, {\rm kg\,
day}$ respectively. The curves of constant $N_{\rm expt}$
are, for $m_{\chi} = 50 \, {\rm GeV}$,  roughly parallel to the 
$m_{\chi}$ axis, improving slightly the
accuracy with which $\sigma_{\rm p}$ can be measured.

For massive WIMPs, $m_{\chi} \gg m_{\rm A}$, the weak dependence of
$E_{\rm R}$ on the WIMP mass means that there is a large spread in the
distribution of ML masses even for large exposures.  There is a small, but
significant, probability that an experiment will happen to observe one or
more unusually large energy events and hence find a large ($>{\cal O}({\rm
  TeV})$) ML mass. Because of the extremely weak dependence of the
energy spectrum on the WIMP mass, the extended likelihood function
also varies weakly with the mass in these cases (where the ML mass is
of order a few TeV or greater, varying the WIMP mass by a factor of
two from the ML value only changes the 4th significant figure of the
log of the extended likelihood function). It is therefore not be
possibly to determine the WIMP mass accurately in these cases.

For an
underlying WIMP mass of $m_{\chi}=250 \, {\rm GeV}$ and an exposure of ${\cal
E}= 3 \times 10^{2} \, {\rm kg \, day}$ it will be difficult to place
any meaningful constraints on the WIMP parameters. As the exposure is
increased it will be possible to place a lower limit on the
mass: for ${\cal E}= 3 \times 10^{3}, \, 3 \times 10^{4}$ and $3
\times 10^{5} \, {\rm kg\, day}$, $95\%$ of experiments have ML
mass greater than $50, 125$ and $200 \, {\rm GeV}$
respectively. However it will not be possible to place an
upper limit tighter than $m_{\chi} < {\cal O}(1\, {\rm TeV})$ 
on the WIMP mass with a reasonable ($>68\%$) degree of confidence. The shape
of the curves of constant $N_{\rm exp}$ for $m_{\chi} \gg m_{\rm A}$ mean
that it will also be difficult to constrain $\sigma_{\rm p}$. 
For ${\cal E}= 3 \times 10^{4}$ and $3 \times 10^{5} \, {\rm kg} $, $95\%
$ of experiments lie within $\Delta (\log{\sigma_{\rm p}})
 \approx {}^{+0.4}_{-0.2}$ and ${}^{+0.4}_{-0.1}$ of the input value of 
$\sigma_{\rm p}$
(the asymmetric spread arises from the asymmetry in the distribution
of ML masses).

The situation
is even worse for an underlying WIMP mass of  $m_{\chi}=500 \, {\rm GeV}$.
For ${\cal E}= 3 \times 10^{2}\, {\rm kg\, day}$ there is a $> 50 \%$
probability that an experiment
will observe two or fewer events. In these circumstances rather than attempting
to determine the WIMP mass and cross-section it would be more reasonable to
instead determine the regions of WIMP mass-cross-section excluded.
The observed number of events with the greatest probability is $N_{\rm exp}=2$.
In the relevant panel of Fig.~\ref{m500} we therefore plot the exclusion limits
from the $68\%$ and $95\%$ upper limits on the underlying mean
number of events if this number of events were observed, $\lambda < 3.5$ and $6.3$ respectively.
For ${\cal E}=3 \times 10^{4}$ and $3
\times 10^{5} \, {\rm kg\, day}$, $95\%$ of experiments have ML
mass greater than $125$ and $350 \, {\rm GeV}$ respectively, however it will
not be possible to place an upper limit on the WIMP mass at more
than $68\%$ confidence. Similarly it will
only be possible to place a lower limit on $\sigma_{\rm p}$.

\begin{figure}  
\begin{center}  
\epsfxsize=6.in  
\epsfbox{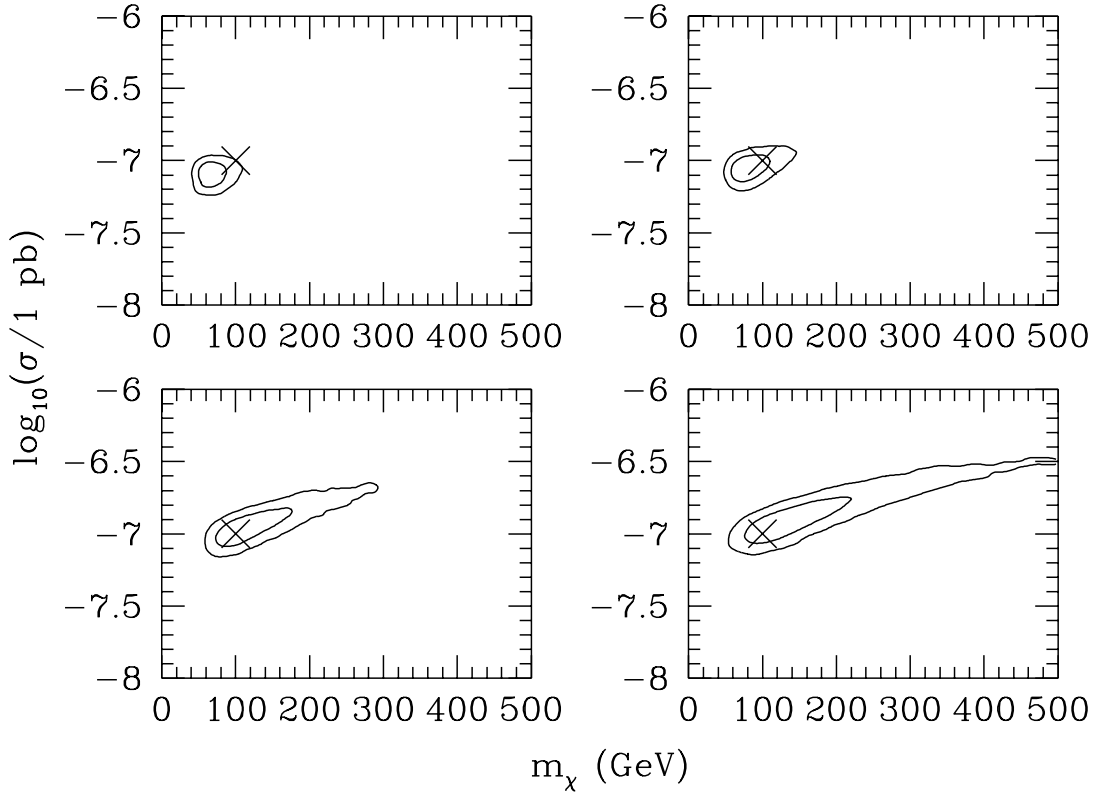} 
\end{center}  
\begin{center} 
\caption{As fig.~\ref{m100} but varying the input circular velocity
$v_{\rm c} = 180, \, 200,\, 240$ and $260 \, {\rm km \, s}^{-1}$ with the 
exposure fixed at ${\cal E}=3 \times 10^{3} \, {\rm kg \, day}$. The maximum likelihood
analysis is carried out assuming $v_{\rm c}= 220 \, {\rm km \, s}^{-1}$.
\label{m100vc}}  
\end{center} 
\end{figure}

\begin{figure}  
\begin{center}  
\epsfxsize=6.in  
\epsfbox{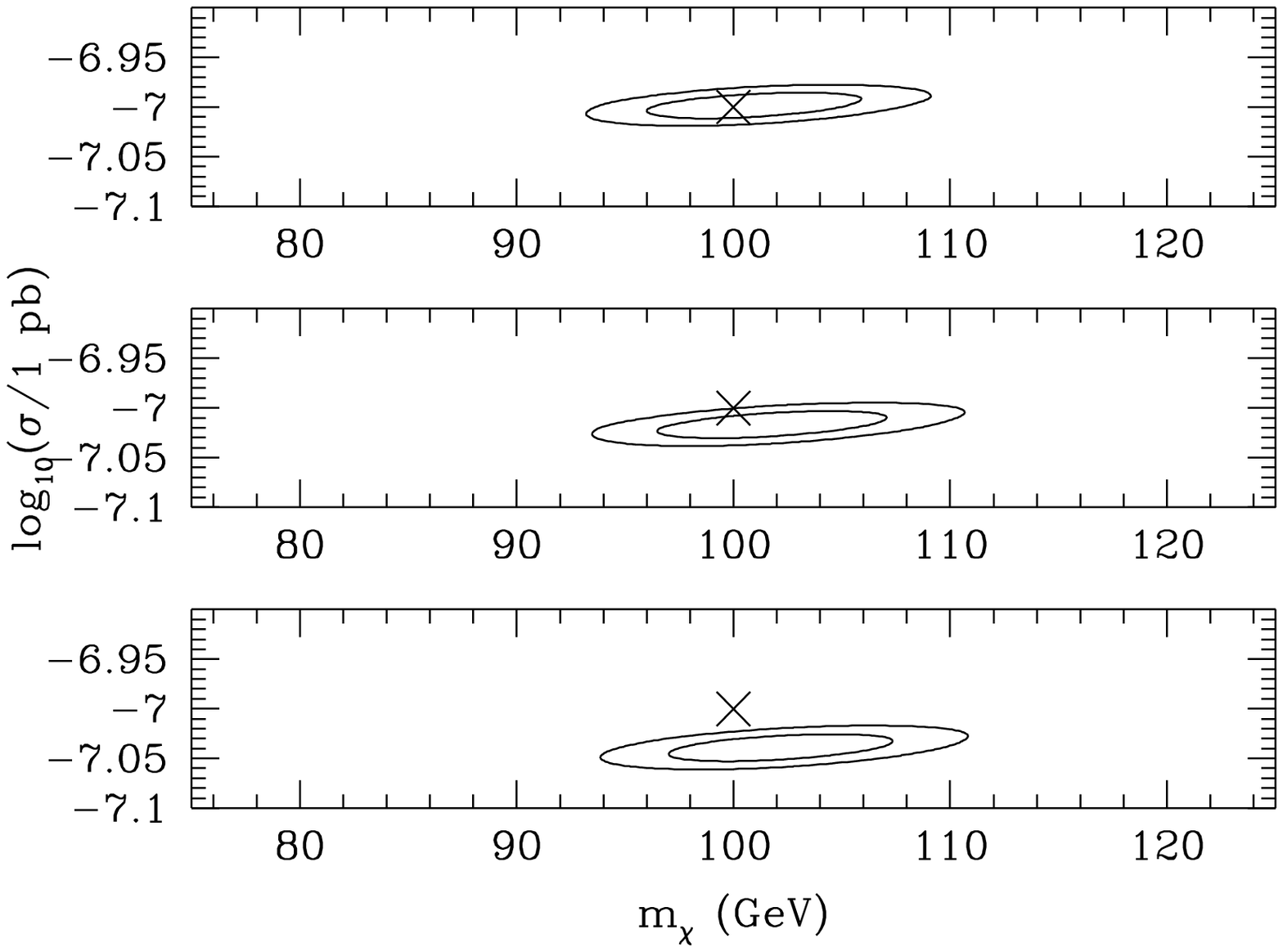} 
\end{center}  
\begin{center} 
\caption{As fig.~\ref{m100vc} but for the standard halo model (top panel)
and the two non-standard halo
models described in the text with the exposure fixed at
${\cal E}=3 \times 10^{5} \, {\rm kg \, day}$. The maximum likelihood
analysis is carried out assuming the standard Maxwellian speed
distribution.
\label{m100ns}}  
\end{center} 
\end{figure}

In fig.~\ref{m100vc} we show the effect of the uncertainty in the
value of the local circular speed, $v_{\rm c}$, on the
determination of $m_{\chi}$. For an input WIMP mass of $m_{\chi}=100\,
{\rm GeV}$ and an exposure of ${\cal E}= 3 \times 10^{3} \, {\rm kg \,
day}$ we vary 
$v_{\rm c}$ between $180$ and $260 \, {\rm km \, s}^{-1}$~\cite{klb}.
The likelihood analysis is however carried out assuming $v_{\rm c}=220
\, {\rm km \, s}^{-1}$. We see in fig.~\ref{m100vc} that there
is a degeneracy between $m_{\chi}$ and $v_{\rm c}$.
The kinetic energies of the incoming WIMPs depend on their mass and 
velocities. For larger (smaller) $v_{\rm c}$ the incoming
WIMPs have larger (smaller) mean kinetic energies than assumed, resulting
in larger (smaller) ML mass values.
This statement can be made more quantitative by differentiating the 
expression for the characteristic energy
$E_{\rm R}$,~eq.~(\ref{er}):
\begin{equation}
    \frac{{\Delta} m_{\chi}}{m_{\chi}} = - [1+ (m_{\chi}/m_{\rm A})]
     \frac{\Delta v_{\rm c}}{v_{\rm c}}  \,.
\end{equation}
This gives, for an input WIMP mass of $m_{\chi}=100\, {\rm GeV}$ a
$ \sim 20 \, {\rm GeV}$ shift in the WIMP mass from a $20 \, {\rm km \,
s}^{-1}$ uncertainty in $v_{\rm c}$. However, as we see in fig.~\ref{m100vc},
the shape of the distribution of the ML parameters changes, in 
qualitatively the same way as when the underlying WIMP mass is changed.
As $v_{\rm c}$ is decreased (increased) the expected number
of events increases (decreases) and hence the best-fit cross-sections
typically decrease (increase).

Finally in fig.~\ref{m100ns} we examine the effect of the uncertainty
in the detailed shape of the local velocity distribution. For an input
WIMP mass of $m_{\chi}=100\, {\rm GeV}$ and the largest exposure, ${\cal E}=
3 \times 10^{5} \, {\rm kg \, day}$, we use as input the logarithmic
ellipsoidal model (which is the simplest triaxial generalization of
the standard isothermal sphere)~\cite{lge} and two of the sets of
parameter values previously considered in Ref.~\cite{lgegreen}:
$p=0.9, \, q=0.8, \, \gamma=0.07$ and $p=0.72, \, q=0.7, \, \gamma=4.0$
with the Earth located on the intermediate axis. These parameters
correspond to axial ratios $1:0.78:0.48$ and $1:0.45:0.38$
(i.e. quite extreme triaxiality, especially in the second case) and, in both 
cases, velocity anisotropy $\beta=0.1$. The local circular and escape speeds
are kept fixed at the values used for the standard halo model.
The likelihood analysis is carried out assuming the standard
Maxwellian velocity distribution. Even for the second, quite extreme,
model the shift in the typical ML mass from the underlying value
is relatively small (less than $5\, {\rm GeV}$). The small increase
in the expected number of events for these halo models leads
to a small downwards shift in the distribution of best-fit cross-sections.

\section{Validity of assumptions}
\label{assumpt}

\subsection{WIMP distribution}

Since the differential event rate is directly proportional to $\rho_{\chi} \,
\sigma_{\rm p}$, any uncertainty in the local WIMP density leads
straightforwardly to an equivalent uncertainty in $\sigma_{\rm p}$. We
have fixed the local WIMP density to the `standard value' of $\rho_{\chi}=0.3
\, {\rm GeV} \, {\rm cm}^{-3}$. Refs.~\cite{ggt,bub} found, using
various observations to constrain the parameters of a range of halo
models, local densities in the range $0.2-0.8 \, {\rm GeV \,
cm}^{-3}$, which would lead to a factor of a few uncertainty in the
determination of the cross-section.

We saw above that an uncertainty in the local circular velocity
translates directly into an uncertainty in the measured WIMP mass.
While the annual modulation signal depends sensitively on the
WIMP velocity distribution~\cite{amnonstand}, the {\em time averaged} differential event 
rates produced by smooth halo models are similar
to that found under the standard assumption of a Maxwellian velocity
distribution (e.g.~\cite{drdens,lgegreen}). Hence, as we saw above,
the uncertainty in the speed distribution, for fixed $v_{\rm c}$, only
leads to a small systematic error in the determination of the WIMP
mass (see fig.~\ref{m100ns}). In other words, for smooth halo models
the characteristic energy of the energy spectrum, and hence
the WIMP mass determination, depends only weakly on the detailed kinetic
energy distribution.  These
smooth halo models are derived by solving the collisionless Boltzmann
equation which assumes that the dark matter distribution has reached a
steady state.  The assumption that the local dark matter distribution
is dynamically mixed, and hence smooth, is, however, a strong
assumption, which may not be valid~\cite{moore,sw}. Helmi, White \&
Springel~\cite{hws} found that the 'particles' in a
simulated Milky-Way like halo in a $\sim (\rm kpc)^3$ volume around
the solar radius were relatively smoothly distributed.  
Direct detection experiments, however, probe the
dark matter distribution on sub-milli-pc scales. The highest
resolution simulations carried out to date have resolution of order
${\cal O} (100 \, {\rm pc})$ and hence can not resolve the dark matter
distribution on the relevant scales.

In CDM cosmologies
structure forms hierarchically and the local dark matter distribution
will depend on the fate of the first, smallest, WIMP micro-halos to form.
The mass of these microhalos is set by the WIMP microphysics in the
early Universe~\cite{wimpmicro}, specifically kinetic decoupling and
free-streaming, and (depending on the WIMP's interaction properties)
is expected to lie in the range $10^{-12} M_{\odot}$ to $10^{-4}
M_{\odot}$~\cite{micromass}. The dynamical evolution of these
micro-halos is being studied~\cite{microdyn}, however the detailed dark
matter phase space distribution on sub-milli-pc scales is not yet
known with any degree of certainty. If the local dark matter
distribution consists of a small number of streams, with a-priori
unknown velocities, then the energy spectrum would consist of a series
of (sloping due to the energy dependence of the form factor) steps. 
The positions of these steps would depend on the WIMP mass and the (unknown)
stream velocities, while the height of the steps would depend on the
(unknown) stream density. In this case it would therefore not be possible
to determine the WIMP mass from the energy spectrum.

\subsection{Negligible background} 

From an experimental point of view, the most significant assumption is
probably that of negligible background. Non-zero background could be
incorporated (c.f. Ref.~\cite{material}) by simulating the recoil
spectra produced by neutrons and including the background event rate
(and additional parameters modeling the spectrum of the background
events) in the maximum likelihood analysis.  This would, however,
require detailed modeling of the detector set-up and shielding.
Adding additional parameters to the maximum likelihood analysis would
clearly degrade the accuracy with which the WIMP mass and
cross-section could be determined. The extent of the degradation would
depend on the shape of the background energy spectrum (and how well it
is known); the more similar it is to the WIMP spectrum the larger
the errors will be.

\subsection{Other sources of systematic error}

Finite energy resolution and uncertainty in the form factor are other
potential experimental sources of systematic error.  The Helm form
factor, with parameter values as advocated by Lewin and
Smith~\cite{ls}, deviates by only of order $1\%$ from that calculated
using electron elastic scattering data~\cite{duda}.  We have checked
that Gaussian energy resolution, with full width at half maximum of
order $1 \, {\rm keV}$~\cite{SuperCDMS1}, does not 
affect the WIMP parameters extracted from the energy
spectrum~\footnote{Since the underlying differential event rate is,
  modulo the form factor, close to exponential, Gaussian smoothing
 only changes its shape for energies of order the resolution (which are below
the energy threshold) .}. Both of these issues are
therefore likely to be less important than non-zero background and/or
fine-grained structure in the WIMP distribution.

From a theoretical point of view the WIMP may have spin-dependent
interactions~\footnote{Natural Germanium contains $7.7\%$ ${}^{73} 
{\rm Ge}$ which
is sensitive to spin dependent interactions~\cite{cdmssd}.}
 with the nucleon and/or different coupling to the
proton and neutron (e.g. Ref.~\cite{wimpsd,cdmssd}). 
The measurement of the WIMP mass,
in principle, in this case has been considered in Ref.~\cite{bk}.

\section{Summary} 
\label{discuss} 

We have examined the accuracy with which it will be possible to
determine the WIMP mass from the energy spectrum observed in a
SuperCDMS-like direct detection experiment given optimistic 
assumptions about the WIMP distribution and detector properties.
If the WIMP distribution  is smooth, the differential event
rate varies with energy, modulo the energy dependence of the detector
form factor, as ${\rm d} R/ {\rm d} E \propto \exp{(-E/E_{\rm R})}$ where
the characteristic energy, $E_{\rm R}$, depends on the WIMP mass,
$m_{\chi}$. For light WIMPs ($m_{\chi} \ll m_{\rm A}$ where $m_{\rm
A}$ is the mass of the target nuclei) $E_{\rm R} \propto m_{\chi}^2$,
while for heavy WIMPs ($m_{\chi} \gg m_{\rm A}$) $E_{\rm R} \sim {\rm
const}$. Consequently for light WIMPs the energy spectrum is strongly
dependent on the WIMP mass, allowing the mass to be measured
fairly accurately. For heavy WIMPs the dependence on the WIMP
mass is far weaker making it difficult to measure the mass.

We have carried out Monte Carlo simulations of a SuperCDMS-like
experiment composed of a ${\rm Ge}$ target with an energy threshold of
$10 \, {\rm keV}$ and zero background. We assumed, initially, that the
local WIMP density is known and that the WIMP speed distribution is
Maxwellian with local circular velocity $v_{\rm c}=220 \, {\rm km \,
  s}^{-1}$. For an optimistic interaction cross-section of $\sigma_{\rm
  p}=10^{-7} \, {\rm pb}$, just below the current exclusion limits
from the CDMS experiment~\cite{CDMS1}, we considered a range of WIMP
masses, $25 \, {\rm GeV} < m_{\chi} < 500 \, {\rm GeV}$, and,
efficiency weighted, exposures, $3 \times 10^{2} \, {\rm kg
  \, day} < {\cal E} < 3 \times 10^{5} \, {\rm kg \, day}$. For ${\cal
  E}= 3 \times 10^{2} \, {\rm kg \, day}$ the expected number of
events is small, the ML masses and cross-sections are stratified
on curves of constant number of events and it would not be possible
to obtain better than order of magnitude constraints on the WIMP
parameters. For ${\cal E} = 3 \times 10^{5} \, {\rm kg \, day}$ and an
input mass of $m_{\chi}=100 \, {\rm GeV}$ it would be possible, given
the validity of the assumptions stated above, to measure the WIMP mass
with accuracy $\pm 10 \, {\rm GeV}$ and the fractional cross-section
with accuracy $\Delta(\log{\sigma_{\rm p}})= \pm 0.02$.  
The mass of lighter WIMPs could be measured more accurately, however 
for very light 
WIMPs, $m_{\chi} <  {\cal O}(10 \, {\rm GeV})$, the number of 
events above the 
detector energy threshold would be too small to allow the mass to be 
measured accurately. For more massive WIMPs
there is a significant tail of experiments with ML masses
significantly larger than the input WIMP mass.  For heavy WIMPs,
$m_{\chi} > {\cal O} (500 \, {\rm GeV})$, even with ${\cal E} = 3
\times 10^{5} \, {\rm kg \, day}$ it will only be possibly to place
lower limits on the WIMP mass and cross-section.

We then examined the effect of varying the underlying WIMP speed
distribution for an input WIMP mass of $100 \, {\rm GeV}$. A change of
$\pm 20 \, {\rm km \, s}^{-1}$ in the local circular speed, $v_{\rm
c}$, leads to a shift in the distribution of best fit WIMP mass of
roughly $\pm 20 \, {\rm GeV}$ (although an increase in $v_{\rm c}$
leads, like an increase in the underlying WIMP mass, to a large tail of
experiments with large best fit masses). Changing the shape of the WIMP
velocity distribution, while keeping $v_{\rm c}$ fixed, leads to only
a small change in the input energy spectrum and hence the shift in the
best fit WIMP masses is relatively small, $< {\cal O}( 5 \, {\rm GeV})$, even
for quite extreme smooth halo models. There is, for smooth halo
models, a factor of a few uncertainty in the local WIMP
density~\cite{ggt,bub} which leads to a corresponding uncertainty in
$\sigma_{\rm p}$. The assumption of a smooth WIMP distribution may
well, however, not be valid. The local WIMP distribution, on sub
milli-pc scales, may be composed of a (a priori unknown) number of
discrete streams with unknown velocities. If this is the case it will
not be possible to extract constraints on the WIMP mass from an observed
signal.

Finally we discussed the validity of the other assumptions made,
namely negligible background, purely spin independent coupling, perfect
energy resolution and known detector form factor. Of these negligible
background is probably the most significant and
could in principle be taken into account by simulating the recoil spectra
produced by neutrons and including the background event rate and energy spectrum
in the maximum likelihood
analysis.

\ack
 AMG is supported by PPARC and is grateful to Ben Morgan and Simon Goodwin
for useful discussions and Meghan Gray and Chris Conselice for assistance
with supermongo contour plotting.

\end{document}